\documentclass[prl,twocolumn,superscriptaddress,nofootinbib]{revtex4}
\usepackage{graphicx}
\usepackage{bm}
\usepackage{amssymb}
\usepackage{color}
\usepackage{amsmath}
\usepackage{mathrsfs} 
\usepackage{amstext}
\usepackage[polish,english]{babel}
\usepackage{latexsym}
\usepackage{array}             
\usepackage{multirow}         
\usepackage[usenames,dvipsnames]{xcolor}
\usepackage[colorlinks=true,citecolor=Blue,linkcolor=RubineRed,urlcolor=Blue]{hyperref}
\usepackage{tocloft}
\usepackage{multibib}
\usepackage{pdfpages}

\newcommand{\pac}[1]{ \left\{ #1 \right\} }
\newcommand{\pap}[1]{\left( #1 \right)}
\newcommand{\pas}[1]{\left[#1 \right]}

\def\la{\langle}
\def\ra{\rangle}

\newcommand{\beq}{\begin{equation}}
\newcommand{\eeq}{\end{equation}}
\newcommand{\beqa}{\begin{eqnarray}}
\newcommand{\eeqa}{\end{eqnarray}}

\begin{document}
\title{Full Counting Statistics of Topological Defects After Crossing a Phase Transition}
\author{Fernando J. G\'omez-Ruiz\href{https://orcid.org/0000-0002-1855-0671}{\includegraphics[scale=0.45]{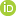}}}
\affiliation{Donostia International Physics Center,  E-20018 San Sebasti\'an, Spain}
\author{Jack J. Mayo}
\affiliation{Donostia International Physics Center,  E-20018 San Sebasti\'an, Spain}
\affiliation{University of Groningen, 9712 CP Groningen, Netherlands}
\author{Adolfo del Campo\href{https://orcid.org/0000-0003-2219-2851}{\includegraphics[scale=0.45]{orcid}}}
\affiliation{Donostia International Physics Center,  E-20018 San Sebasti\'an, Spain}
\affiliation{IKERBASQUE, Basque Foundation for Science, E-48013 Bilbao, Spain}
\affiliation{Department of Physics, University of Massachusetts, Boston, MA 02125, USA}
\begin{abstract}
We consider the number distribution of topological defects resulting from the finite-time crossing of a continuous phase transition  and identify signatures of universality beyond the mean value, predicted by the Kibble-Zurek mechanism. Statistics of defects follows a binomial distribution with $\mathcal{N}$ Bernouilli trials associated with the probability of forming a topological defect at the locations where multiple domains merge. All cumulants of the  distribution are  predicted to exhibit a common universal power-law scaling with the quench time in which the transition is crossed. Knowledge of the distribution is used to discuss the onset of adiabatic dynamics and bound rare events associated with large deviations.\\
\\
DOI: \href{https://journals.aps.org/prl/abstract/10.1103/PhysRevLett.124.240602}{10.1103/PhysRevLett.124.240602}
\end{abstract}
\maketitle

In a scenario of spontaneous symmetry breaking, the dynamics of a system across a  continuous phase transition is described by the Kibble-Zurek mechanism (KZM)~\cite{Kibble76a,Kibble76b,Zurek96a,Zurek96c}. When the transition is driven in a finite quench time $\tau_Q$,  KZM predicts the formation of domains of volume $\hat{\xi}^D$, where $D$ is the spatial dimension of the system. Specifically,  KZM uses as input  the equilibrium value of the correlation length $\xi$ and the relaxation time $\tau$. By varying a control parameter $\lambda$ across the critical value $\lambda_c$ both quantities exhibit a power-law divergence as a function of the distance to the critical point $\epsilon=(\lambda-\lambda_c)/\lambda_c$,
\beqa
\xi=\xi_0|\epsilon|^{-\nu},\quad \tau=\tau_0|\epsilon|^{-z\nu}.
\eeqa
Here, $\nu$ is the  correlation-length critical exponent and $z$ denotes the dynamic critical exponent. Both are determined by the universality class of the system. By contrast, $\xi_0$ and $\tau_0$ are microscopic constants.
 KZM states that when the phase transition is driven in a time scale $\tau_Q$ by a linear quench of the form $\epsilon=t/\tau_Q$, domains in the broken symmetry phase spread over a length scale
\beqa
\hat{\xi}=\xi_0\left(\frac{\tau_Q}{\tau_0}\right)^{\frac{\nu}{1+z\nu}}.
\eeqa
In $D$ spatial dimensions, KZM predicts the mean number of topological defects to scale as
\beqa
\la n\ra\propto\left(\frac{\tau_0}{\tau_Q}\right)^{\frac{D\nu}{1+z\nu}}.
\eeqa
This power law behavior with the quench time, initially derived for classical systems, similarly describes the dynamics across a quantum phase transition~\cite{Dziarmaga10,Polkovnikov11,DZ14}. In this context, the scaling is generally studied in the residual mean energy and the number of quasi-particles, which generally differs from the number of topological defects~\cite{Uwe07,Uwe10,Uhlmann10}.  
The KZM has also been extended to a variety of scenarios including nonlinear quenches~\cite{Diptiman08,Barankov08,Fernando19}, long-range interactions~\cite{Caneva08,Acevedo14,Hwang15,Lukin17,Defenu18,Puebla19}, and inhomogeneous phase transitions in both classical~\cite{KV97,ZD08,Zurek09,delcampo10,DRP11,DKZ13,DZ14} and quantum systems~\cite{DM10,DM10b,CK10,Rams16,Nishimori18,Fernando19}.  KZM has been experimentally investigated in a wide variety of platforms reviewed in \cite{DZ14}, with recent tests being performed in trapped ions~\cite{EH13,Ulm13,Pyka13}, colloidal monolayers~\cite{Keim15}, ultracold Bose and Fermi gases~\cite{Weiler08,Lamporesi13,Chomaz15,Navon15,Shin19}, and quantum simulators~\cite{Guo14,Wang14,Wu16,Cui16,Lukin17}. 

Despite this progress,  features of the counting statistics of defects  other than the mean number have received scarce attention. An exception concerns scenarios of ${\rm U}\pap{1}$ symmetry breaking leading to, e.g., the spontaneous current formation in  a superfluid confined in a toroidal trap or a  superconducting ring~\cite{Zurek96a,Zurek96c,Monaco02,Das12,Sonner15,Nigmatullin16}. 
While the average circulation vanishes, it was shown that its variance is consistent with a one-dimensional random walk model in which the number of steps is predicted by the circumference of the ring divided by the KZM length scale $\hat{\xi}$~\cite{Zurek96a,Zurek96c}. It is however not clear how to extend this argument to higher dimensions~\cite{Uwe07}.  Not long ago, the distribution of kinks formed in a quantum Ising chain driven from the paramagnetic to the ferromagnetic phase was studied both theoretically~\cite{delcampo18} and in the laboratory~\cite{Cui19,Bando20}.

In this letter, we focus on signatures of universality beyond the mean number of topological defects and show that the full counting statistics of topological defects is actually universal. In particular, we argue that  {\it i)} the defect number distribution is binomial, {\it ii)} all cumulants are proportional to the mean and scale as a universal power law with the quench rate {\it iii)}  this power law is fixed by the conventional KZM scaling. This knowledge allows us to characterize universal features regarding the onset of  adiabatic dynamics (probability for no defects) and  deviations of the number of kinks away from the mean value.

{\it Number distribution of topological defects.---} To estimate the defect number distribution we assume that the number of domains in the total system is set by
\beqa
\mathcal{N}_D=\frac{{\rm Vol}}{\hat{\xi}^D},
\eeqa
where ${\rm Vol}$ denotes the volume of the system. Topological defects may form at the interface between multiple domains. For instance, the formation of vortices has been demonstrated  by merging independent Bose-Einstein condensates~\cite{Scherer07}. The same principle is at the core of phase-imprinting methods for soliton formation~\cite{Burger99}. Disregarding boundary effects, the number of locations where a topological effect may be formed is approximately given by $\mathcal{N}=\mathcal{N}_D/f$ where $f$ takes into account the average   number of domains that meet at a point. Alternatively, $f$ can be considered a fudge factor.

We next propose that at the merging between multiple domains a topological defect forms with a  probability $p$. Similarly, no topological defect will be formed at any such location  with probability $1-p$. The formation of topological defects at different locations is assumed to be independent  and in each case the event of formation can be associated with a Bernouilli random variable. We thus propose that the number distribution of topological defects can be  approximated by the the binomial distribution with parameters $\mathcal{N}$ and $p$. This is the discrete probability distribution for the number of successes (number of topological defects formed) in a sequence of $\mathcal{N}$ independent trials:
\beqa
P(n)\sim B(n,\mathcal{N},p)= \begin{pmatrix}
\mathcal{N}  \\
n
\end{pmatrix}  p^n\,(1-p)^{\mathcal{N}-n}.
\eeqa
Thus, $P(n)$ is centered at 
\beqa
\label{meankzm}
\la n\ra=\frac{p{\rm Vol}}{f\hat{\xi}^D}=\frac{p{\rm Vol}}{f\xi_0^D}\left(\frac{\tau_0}{\tau_Q}\right)^{\frac{D\nu}{1+z\nu}},
\eeqa
in agreement with the KZM scaling. Further, its  variance is set by
\beqa
\label{varbkzm}
{\rm Var}(n)=\la n^2\ra-\la n\ra^2=\frac{{\rm Vol}}{f\hat{\xi}^D}p(1-p)\propto\tau_Q^{-\frac{D\nu}{1+z\nu}},
\eeqa
and  is always proportional to the mean, as ${\rm Var}(n)=(1-p)\la n\ra$.

{\it High-order cumulants.---} To further characterize the number distribution of defects it is convenient to introduce the Fourier transform $\widetilde{P}(\theta)$ of $P(n)$, satisfying \cite{Cramer46} %
\beqa
\label{pneq}
P(n)=\frac{1}{2\pi}\int_{-\pi}^{\pi}d\theta \;\widetilde{P}\pap{\theta} \exp\pas{-i\theta n},
\eeqa
and  known as  the characteristic function, $\widetilde{P}(\theta)=\mathbb{E}[e^{i\theta n}]$. Its logarithm is the cumulant generating function. Specifically,  cumulants $\kappa_q$ of $P(n)$ are defined using the expansion
\beqa
\log  \widetilde{P}(\theta)=\sum_{q=1}^\infty \frac{(i\theta)^q}{q!}\kappa_q.
\eeqa
For the binomial distribution the cumulant generating function reads
\beqa
\log\tilde{P}(\theta)=\la n\ra\log(1-p+pe^{i\theta})
\eeqa
whence it follows that all cumulants are proportional to the mean and thus scale universally with the quench time,
\beqa
\label{kappaqpl}
\kappa_q\propto \left(\frac{\tau_0}{\tau_Q}\right)^{\frac{D\nu}{1+z\nu}}.
\eeqa
They satisfy the recursion relation $\kappa_{q+1}=p(1-p)d\kappa_q/dp$  and those with $q>2$ signal non-normal features of the distribution. For instance,  $\kappa_3/\la n\ra=p(1-p)(1-2p)$ and  $\kappa_4/\la n\ra=p(1-p)(1-6p+6p^2)$.

However, it follows from central limit (De Moivre-Laplace) theorem that for large $\mathcal{N}$ with $p$ constant the distribution becomes asymptotically normal \cite{Cramer46}, i.e., 
\beqa\label{NormPn}
P(n)\sim\frac{1}{\sqrt{2\pi(1-p)\la n\ra}} \exp\pas{-\frac{\pap{n-\la n\ra}^2}{2(1-p)\la n\ra}}, 
\eeqa

where $\la n\ra$ is given by~\eqref{meankzm} in agreement with KZM and we have used that the variance is proportional to the mean, according to Eq.~\eqref{varbkzm}.

{\it Non-uniform probabilities for defect formation.---} We have assumed at the interface between multiple domains topological defects form with constant probability $p$. One can generally expect this not to be the case. For instance, according to the geodesic rule the  probability for defect formation depends on the number of domains that merge at the location of interest~\cite{Kibble76a,Chuang91,Bowick94,Scherer07}. One may wonder how the defect number distribution is affected when the probability for formation of topological defect is not fixed but varies at different locations. Keeping the assumption that the events of formation of topological defects are independent, the number of defects formed is thus given by the sum of independent Bernouilli trials, in which the probabilities for defect formation are  $\{p_1,p_2,\dots,p_{\mathcal N}\}$. The resulting distribution is the so called Poisson binomial distribution with characteristic function $\tilde{P}(\theta)=\prod_{j=1}^{{\mathcal N}}(1-p_j+p_je^{i\theta})$ and mean $\la n\ra=\sum_{j=1}^{{\mathcal N}}p_j$ and variance ${\rm Var}(n)=\sum_{j=1}^{{\mathcal N}}p_j(1-p_j)$. This probability distribution actually describes the  distribution of the number of pairs of quasi-particles  in quasi-free fermion models (one dimensional Ising and XY chains, Kitaev model, etc.)~\cite{delcampo18,Cui19}.  Clearly, the mean $ \la n\ra={\mathcal N}\bar{p}$ where the average formation probability $\bar{p}=\sum_{j=1}^{{\mathcal N}}p_j/{\mathcal N}$. Similarly, it is known that ${\rm Var}(n)={\mathcal N}[\bar{p}(1-\bar{p})-s_p^2]$ where $s_p^2=\sum_{j=1}^{{\mathcal N}}(p_j-\bar{p})^2/{\mathcal N}$ is the variance of the distribution $\{p_1,p_2,\dots,p_{\mathcal N}\}$ \cite{Wang93}. Assuming the later to be small, for large ${\mathcal N}$, both ${\rm Var}(n)$ and $\la n\ra$ are proportional to ${\mathcal N}$ and inherit a universal power-law scaling with the quench time.

{\it Onset of  adiabaticity.---} Many applications in statistical mechanics, condensed matter and quantum science and technology require the suppression of topological defects. This is the case in the preparation of novel phases of matter in the ground state or the suppression of errors in classical and quantum annealing. Strict adiabaticity can be associated with the probability to have no defects at all, i.e., $P(0)$.  The latter is given by
\begin{equation}
P\left(0\right)=(1-p)^{\frac{{\rm Vol}}{f\hat{\xi}^D}}\approx \exp\pas{-\frac{{\rm Vol}}{f\hat{\xi}^D}p}=\exp(-\la n\ra),
\end{equation}

where the last term holds for small $p$. In this case, the probability for zero defects decays exponentially with the mean number of defects, i.e., $\la n\ra=\frac{p{\rm Vol}}{f\hat{\xi}^D}$. As a result,
\begin{equation}\label{pzero}
\log\pas{P\pap{0}}=-\frac{p{\rm Vol}}{f\xi_0^D}\left(\frac{\tau_0}{\tau_Q}\right)^{\frac{D\nu}{1+z\nu}},
\end{equation}
a prediction we shall test below.

Relaxed notions of adiabaticity, not based in $P(0)$, can be imposed by considered the cumulative probability in the tails of the distribution, for which  explicit expressions can be found with the binomial model and its normal approximation; see \cite{SM}. It is also possible to find robust bounds, e.g. by considering the tails of the distribution associated with high kink numbers. For  example,  using the Chernoff bound  the upper tail is constrained by the inequality $P\pap{n\geq \la n\ra+\delta}\leq \exp\pas{-\frac{\delta^2}{2\la n\ra +\delta/3}}$~\cite{SM}.

{\it Numerical results.---} For the sake of illustration, we consider  the breaking of parity symmetry in a second-order phase transition~\cite{Laguna98}. Specifically,  we analyze a one-dimensional chain exhibiting a structural phase transition between a linear and a doubly-degenerate zigzag  phase. This scenario is  of relevance to  trapped ion chains~\cite{Retzker08,delcampo10}, confined colloids and dusty plasmas~\cite{Mansoori14}, to name some relevant examples. In the course of the phase transition, parity is broken  and kinks form at the interface between adjacent domains.  To describe the dynamics we consider a lattice description in which each site is endowed with a transverse degree of freedom $\phi_i$ and the total potential reads
\beqa
\label{poteq}
V(\{\phi_i\},t)=\sum_i\frac{1}{2}[\lambda(t)\phi_i^2+\phi_i^4]+c\sum_i\phi_i\phi_{i+1},
\eeqa 
where $\{\phi_i\}$ are real continuous variables and $i=1,\dots,N$. As the coefficient $\lambda(t)$ is ramped from a positive initial value to a negative one, the local single-site potential evolves from a single-well to a double well. The nearest-neighbor coupling favors ferromagnetic order when $c<0$ and antiferromagnetic otherwise. The evolution across the critical point $\lambda_c$ is described by  Langevin dynamics 
\beqa
\label{laneq}
\ddot{\phi}_i+\eta \dot{\phi}_i+\partial_{\phi_i}V(\{\phi_i\},t)+\zeta=0, \,i=1,\dots, N
\eeqa
where $\eta>0$ accounts for friction and $\zeta=\zeta(t)$ is a real Gaussian process with zero mean.  Eqs.~\eqref{poteq} and~\eqref{laneq}  account for the Langevin dynamics of a $\phi^4$-theory on a lattice. This system  is well described by Ginzburg-Landau theory and is characterized by mean-field critical exponents $\nu=1/2$ and $z=2$ in the over-damped regime \cite{Laguna98,delcampo10}. The dynamics is induced by a ramp of $\lambda(t)$ from the value $\lambda(0)=\lambda_0$ to $\lambda(\tau_Q)=\lambda_f$  in the quench time $\tau_Q$ according to $\lambda(t)=\lambda_0+|\lambda_f-\lambda_0|t/\tau_Q$ across the critical point $\lambda_c=2c$, see \cite{SM,Antunes06} for details.
\begin{figure}[t]
\includegraphics[width=1.0\linewidth]{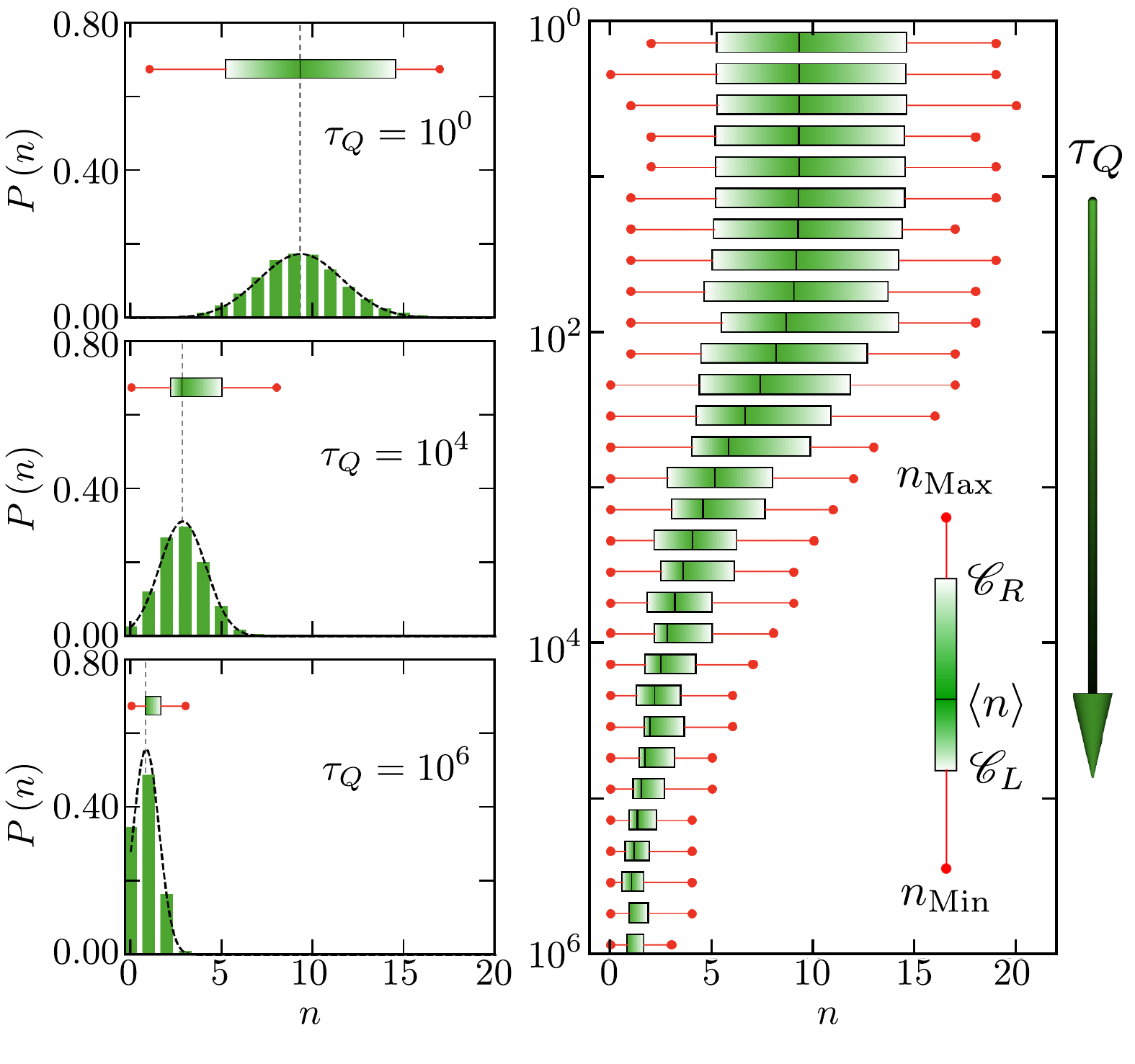}
\caption{\label{fig1}  {\bf Characterization of probability distribution of topological defects.} The left column shows the probability distribution of the number of kinks $P(n)$ generated as a function of the quench time $\tau_Q$.
 The numerical histograms are compared with the normal approximation~\eqref{NormPn} and the dashed vertical line denotes the mean value $\langle n \rangle$. The right panel shows the total distribution of kinks in a box-and-whisker chart, for different quench times and a chain of  $N=100$ sites, using $15000$ trajectories.  
$\mathscr{C}_{R}$ and $\mathscr{C}_{L}$ denote the cumulative probability above and below the mean.  
}
\end{figure}

\begin{figure}[t]
\centering
\includegraphics[width=0.98\linewidth]{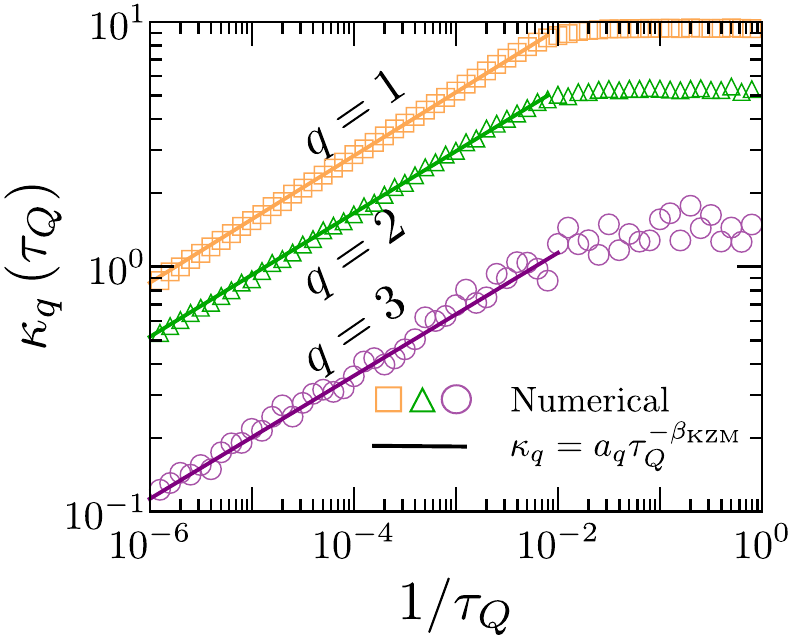}
\caption{\label{fig_kappaq}  {\bf Universal scaling of the cumulants $\kappa_q$ of the kink number distribution.}  From top to bottom, the mean kink density  ($q=1$), its  variance ($q=2$) and the third centered-moment ($q=3$) are shown as a function of the inverse quench time $\tau_Q$  for a chain of  $N=100$ sites and  $15000$ trajectories. Symbols represent numerical data while solid lines describe the analytical approximation derived in the scaling limit, with $\beta_{\rm KZM}=\nu/(1+z\nu)$. 
}
\end{figure}

Full counting statistics of kinks is built by sampling over an ensemble of  $15000$ trajectories; 
see Fig. \ref{fig1} and~\cite{SM} for lower sampling. The mean and width of the distribution are reduced for increasing quench times.
 Histograms for $P(n)$ are shown to be well-reproduced  by the normal approximation~\eqref{NormPn} away from the onset of adiabatic dynamics when the value of $P(0)$ is significant. 
 
 The universal power-law scaling of the cumulants as a function of the quench time is shown in Fig.~\ref{fig_kappaq}. A  fit to the mean number of kinks yields $\kappa_1=(30.838\pm0.297)\tau_Q^{-0.251\pm0.001}$,
 in good agreement with  the KZM, which predicts the power-law exponent $\beta_{\rm KZM}=\nu/(1+z\nu)=1/4$ for mean-field values $\nu=1/2$, $z=2$. Signatures of universality beyond KZM are evident from the scaling of higher order cumulants. Non-normal features of the distribution are signaled by the non-zero value of $\kappa_q$ with $q\geq 3$. The variance  scales as $\kappa_2=(16.948\pm0.217)\tau_Q^{-0.252\pm0.001}$,
  while the third cumulant is fitted to $\kappa_3=(3.621\pm 0.281)\tau_Q^{-0.251\pm 0.011}$.
  Power-law exponents are thus found as well in excellent agreement with the theoretical prediction in Eq.~\eqref{kappaqpl}. 

\begin{figure}[t!]
\centering
\includegraphics[width=1.0\linewidth]{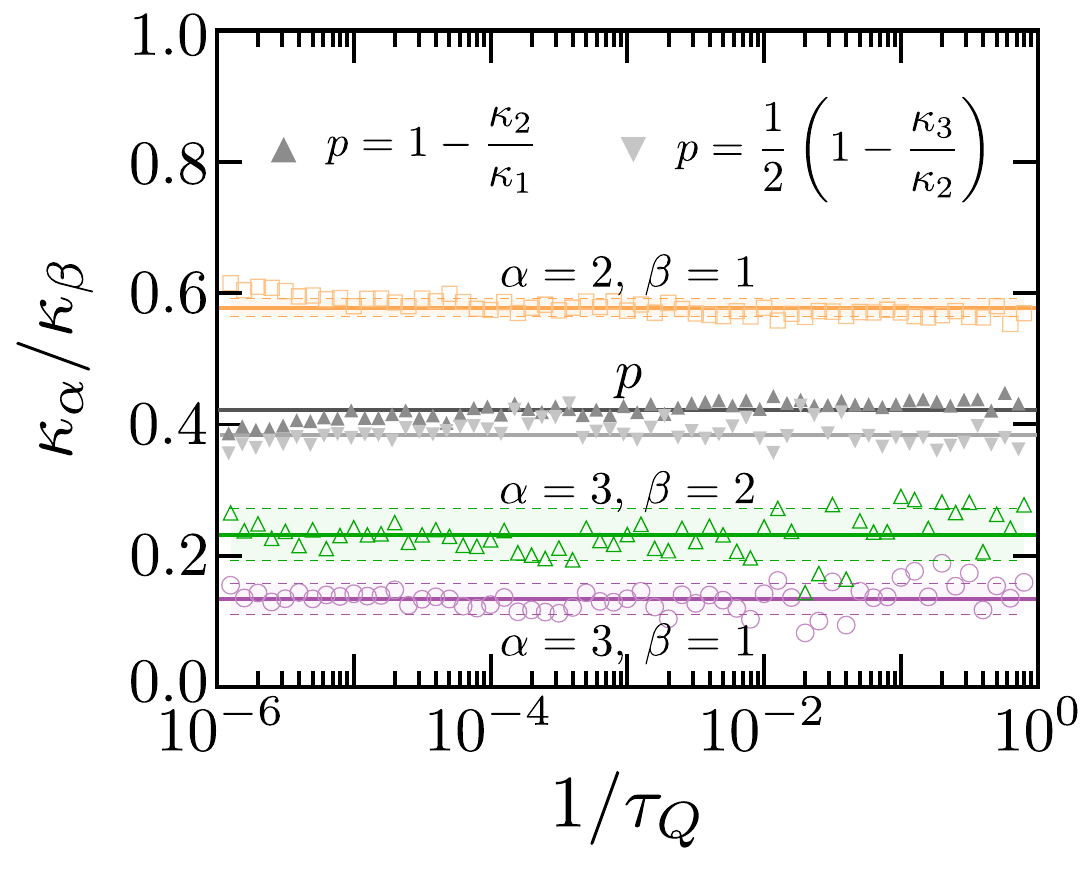}
\caption{\label{fig_3} {\bf Ratio between the first three cumulants as a function of the quench rate.} The numerical results (symbols) for the ratio between the cumulants $\kappa_\alpha$ and  $\kappa_\beta$, where  $\alpha>\beta$ and $\alpha, \beta\in\pac{1,2,3}$, are depicted as function of  the inverse quench time $\tau_Q$. The solid line corresponds to the average of the ratio $\kappa_{\alpha}/\kappa_{\beta}$ and the shadow region between two dashed lines corresponds to the uncertainty associated with each cumulant ratio. Additionally, we showed the numerical (symbols) and mean value (solid lines) of $p$ calculated according to the plot legends.}
\end{figure}

We note however that there is an infinite number of distributions  in which cumulants exhibit a universal scaling with the quench rate of the form $\kappa_q=a_q\tau_Q^{-\beta_{\rm KZM}}$. According to our model for the  full kink counting statistics, the ratio between any two cumulants is independent of the quench time and fixed by the probability $p$ for kink formation at the merging between adjacent domains.  In particular,  $\kappa_2/\kappa_1=1-p$ and  $\kappa_3/\kappa_1=(1-p)(1-2p)$.  Figure~\ref{fig_3} shows the ratio between the first three cumulants as a function of the quench rate. The numerical results are in excellent agreement with the theoretical prediction. In particular, it is found that the observed cumulant ratios $\kappa_2/\kappa_1 = 0.578\pm 0.014$, $\kappa_3/\kappa_1 = 0.134\pm0.023$ and $\kappa_3/\kappa_2=0.232\pm0.040$,
are consistent with a single well-defined value of the probability for kink formation $p=0.422\pm 0.014$;
see~\cite{SM}.

\begin{figure}[h!]
\centering
\includegraphics[width=1\linewidth]{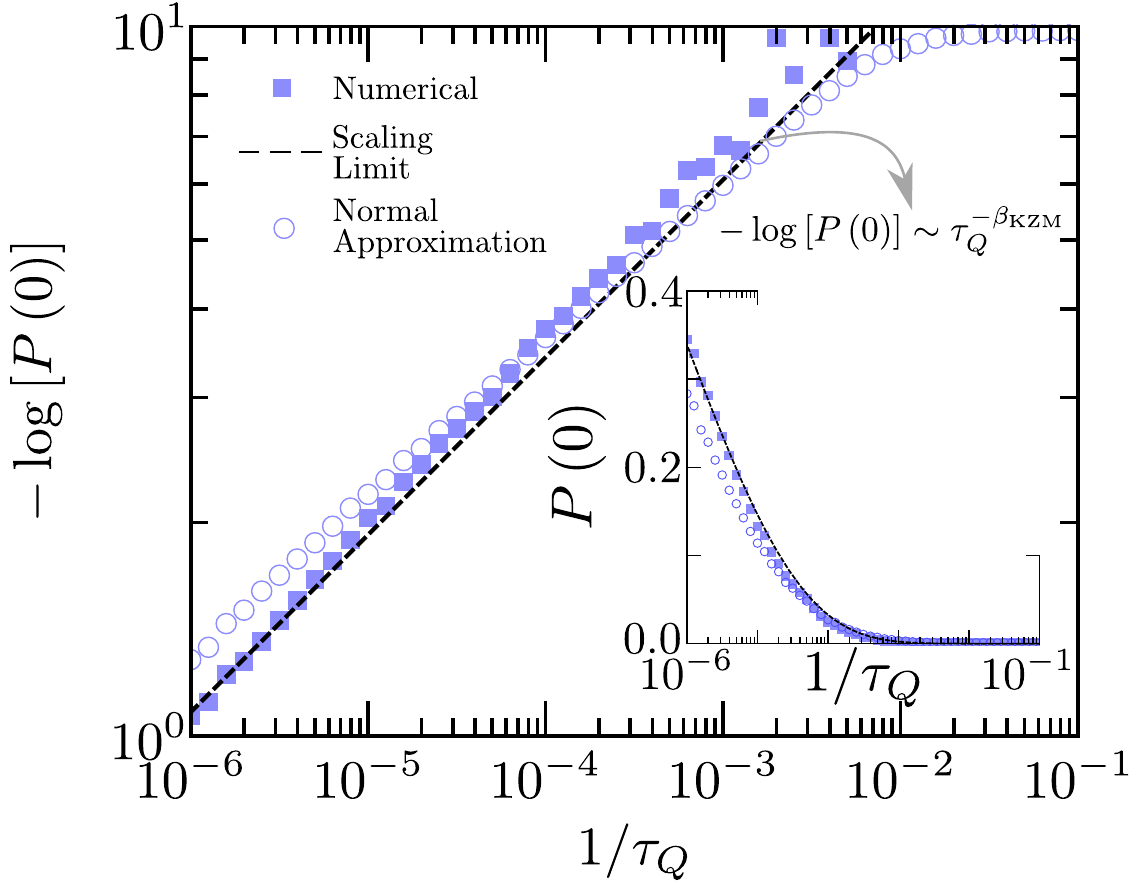}
\caption{\label{fig_4} {\bf Universal scaling of the probability for no kinks $P(0)$ as a function of quench time}. 
 The dashed lines show the universal scaling of the probability for no kinks, as predicted by Eq.~\eqref{pzero}, plotted as a reference with $\beta_{{\rm KZM}}=\nu/\pap{1+z\nu}$. Numerical data (squares) is in excellent agreement with the theoretical prediction. Additionally, using Eq.~\eqref{NormPn} with $n=0$, we show the normal approximation for large  $\mathcal{N}$ (circles) with the fitted value of $p$ in Fig.~\ref{fig_3}.}
\end{figure}

As further evidence for our model, we analyze the probability for no kink formation $P(0)$ as a function of the quench time in Figure~\ref{fig_4}. Its numerical value estimated from the histogram constructed with the ensemble of trajectories follows the theoretical prediction Eq.~\eqref{pzero}.  Thus, Figure~\ref{fig_4} confirms that $P(0)$  decays exponentially with the mean number of kinks, which exhibits itself a universal power-law scaling. At fast quenches, $P(0)$ approaches zero and the comparison is limited by the finite sampling, and the saturation of $\kappa_1$  in  Fig.~\ref{fig_kappaq} due to finite-size effects.
 The normal approximation,  $P(0)=\frac{1}{\sqrt{2\pi(1-p)\la n\ra}}\exp\pas{-\frac{\la n\ra}{2\pap{1-p}}}$, works well for moderate quench rates when $P(n)$ is symmetric  and in absence of finite-size effects,  losing  accuracy at  the onset of adiabaticity, when $P(0)$ is significant. This is shown in Fig.~\ref{fig_4} for the estimated $p=0.422\pm 0.014$ extracted from the mean number of kinks (e.g.  in Fig.~\ref{fig_kappaq}). 
As with $P(0)$, we note that other notions of deviations away from the mean are also shown to be constrained by KZM scaling, as shown in~\cite{SM}.

{\it Summary.---}
When a continuous phase transition is traversed in a finite time scale $\tau_Q$, topological defects form.
The average number  scales with the quench time $\tau_Q$ following a universal power-law scaling predicted by the Kibble-Zurek mechanism.  The same scaling describes the density of excitations  in the quantum domain as well. Given a system whose critical dynamics is described by  KZM, we have argued that the full number distribution of topological defects  is universal and described by a binomial distribution. This model assumes that in the course of the critical dynamics, 
the system size is partitioned in domains of length scale given by the KZM correlation length.  The event of topological defect formation at the interface between multiple domains is associated with a discrete random variable with a fixed success probability.
A testable prediction is that all cumulants of the distribution are proportional to the mean and thus inherit a universal power-law scaling with the quench time, while cumulant ratios are constant and uniquely determined by the probability for kink formation. Other quantities such as the probability for no defects and the deviations away from the mean also exhibit a universal dependence on the quench time. Our findings motivate the quest for universal signatures in the counting statistics of topological defects across the wide variety of experiments used to test KZM dynamics, using e.g.,  convective fluids~\cite{Casado01,Casado06}, colloids~\cite{Keim15}, cold atoms~\cite{Weiler08,Lamporesi13,Chomaz15,Navon15,Shin19}, and trapped ions  \cite{EH13,Ulm13,Pyka13}.

{\it Acknowledgment.--}  
The authors are indebted to  Martin B. Plenio and Alex Retzker for illuminating discussions.
It is also a pleasure to acknowledge discussions with  Micha\lpb{} Bia\lpb{}o\'nczyk, Uwe R. Fischer, Jee Woo Park and Yong-Il Shin, and to thank the Department of Physics at Seoul National University for hospitality.

\bibliography{fcs_defects_Bib}	
\newpage

\pagebreak
\clearpage
\widetext

\begin{center}
\textbf{\large ---Supplemental Material---\\
 Full Counting Statistics of Topological Defects After Crossing a  Phase Transition}\\
\vspace{0.5cm}
Fernando J. G\'omez-Ruiz$^{1}$,  Jack J. Mayo$^{1,2}$ \& Adolfo del Campo$^{1,3,4}$\\
\vspace{0.2cm}
$^1${\it Donostia International Physics Center,  E-20018 San Sebasti\'an, Spain}\\
$^2${\it University of Groningen, 9712 CP Groningen, Netherlands}\\
$^3${\it IKERBASQUE, Basque Foundation for Science, E-48013 Bilbao, Spain}\\
$^4${\it Department of Physics, University of Massachusetts, Boston, MA 02125, USA}
\end{center}
\setcounter{equation}{0}
\setcounter{figure}{0}
\setcounter{table}{0}
\setcounter{section}{0}
\setcounter{page}{1}
\makeatletter
\renewcommand{\theequation}{S\arabic{equation}}
\renewcommand{\thefigure}{S\arabic{figure}}
\renewcommand{\bibnumfmt}[1]{[S#1]}
\renewcommand{\citenumfont}[1]{S#1}
\newcolumntype{M}[1]{>{\centering\arraybackslash}m{#1}}
\begin{center}
\vspace{1.3cm}
{\bf Contents}
\end{center}
\begin{enumerate}
\itemsep0.5em 
\item[\textcolor{RubineRed}{\bf I.}] \textcolor{RubineRed}{\bf Location of the critical point}\hfill\textcolor{RubineRed}{1}
\item[\textcolor{RubineRed}{\bf II.}] \textcolor{RubineRed}{\bf Full counting statistics of kinks as a function sampling}\hfill\textcolor{RubineRed}{2}
\item[\textcolor{RubineRed}{\bf III.}] \textcolor{RubineRed}{\bf Cumulants ratios}
\hfill\textcolor{RubineRed}{4}     
\begin{enumerate}
\item[\textcolor{RubineRed}{A.}] \textcolor{RubineRed}{Numerical estimation of $p$}\hfill\textcolor{RubineRed}{5}
\end{enumerate}
\item[\textcolor{RubineRed}{\bf IV.}] \textcolor{RubineRed}{\bf Onset of  adiabaticity}\hfill\textcolor{RubineRed}{6}
\item[\textcolor{RubineRed}{\bf V.}] \textcolor{RubineRed}{\bf Tails of the number distribution of topological defects}\hfill\textcolor{RubineRed}{6}
\item[] \textcolor{RubineRed}{{\bf References}}\hfill\textcolor{RubineRed}{8} 
\end{enumerate}
\section{I.\quad Location of the critical point}
In what follows we  locate the critical point of the lattice model using a standard approach. 
We focus on the behavior as $\lambda(t) \rightarrow \lambda_{c}$ from a positive initial value $\lambda_{0}>0$. We note that far above the critical point, when $\lambda(t) \gg 1$ and  $\lambda(t) \gg c$ , the system behaves as a set of independent harmonic oscillators. As $\lambda(t)$ is dropped, the contribution of the non-linearity and the nearest-neighbor coupling becomes more relevant.
The equilibrium configuration above the critical point is determined by minimizing the potential according to  $\partial_{\phi_i}V=0$, which yields  $\phi_{i}^{(0)}=0$  for $i=1,\dots,N$. To find the critical value of $\lambda_c$, we consider 
the linearized potential around the equilibrium configuration 
\begin{equation}
V \approx \frac{1}{2}\sum_{i,j}\mathcal{K}_{ij}\phi_{i}\phi_{j},
\end{equation}
where
\begin{equation}
\mathcal{K}_{ij}= \frac{\partial^{2}V}{\partial\phi_{i}\partial\phi_{j}}\bigg\rvert_{\phi_{i,j}=0} = \delta_{i,j}\lambda+c(\delta_{i+1,j}+\delta_{i-1,j}).
\end{equation}
In absence of an environment, the equation of motion of $\phi_{i}$ is
\begin{equation}
\ddot{\phi_{i}}=-\frac{1}{2}\sum_{i',j'}\mathcal{K}_{i'j'}(\delta_{i,i'}\phi_{j'}+\delta_{i,j'}\phi_{i'}),
\end{equation}
which explicitly reads
\begin{equation}
\ddot{\phi_{i}} = -\lambda\phi_{i} - c(\phi_{i+1}+\phi_{i-1}).
\end{equation}
To characterize the normal modes, we use the ansatz
\begin{equation}
\phi_{j}=\sqrt{\frac{2}{N}}\sum_{k>0}(A_{k}\cos(jk)+B_{k}\sin(jk))e^{i\omega_{k}t},
\end{equation}
subject to periodic boundary conditions, $\phi_{0}=\phi_{N}$.
As a result, the coefficients $A_k$ and $B_k$ are related 
\begin{equation}
A_{k}=\frac{B_{k}\sin(Nk)}{1-\cos(Nk)},
\end{equation}
and thus
\begin{equation}
\phi_{j}=\sqrt{\frac{2}{N}}\sum_{k>0}B_{k}\left(\frac{\sin(Nk)}{1-\cos(Nk)}\cos(jk)+\sin(jk)\right)e^{i\omega_{k}t},
\end{equation}
where $k=\frac{2n\pi}{N}$ for $n\in \mathbb{Z}$. The frequencies $\omega_{k}$ are given by
\begin{equation}
\omega_{k}=\sqrt{\lambda+c\frac{\tilde{\phi}_{j-1}(k)+\tilde{\phi}_{j+1}(k)}{\tilde{\phi}_{j}(k)}},
\end{equation}
where
\begin{equation}
\tilde{\phi}_{j}(k) = \frac{\sin(Nk)}{1-\cos(Nk)}\cos(jk)+\sin(jk).
\end{equation}
As $\lambda \rightarrow \lambda_{c}$ from above, the ``soft'' mode driving the transition can be identified as first divergent mode  for which the frequency becomes purely imaginary. Thus the critical value of $\lambda$ is the solution to
\begin{equation}
\label{phimin}
\lambda_{c} = -c\min_{k}\frac{\tilde{\phi}_{j-1}(k)+\tilde{\phi}_{j+1}(k)}{\tilde{\phi}_{j}(k)}.
\end{equation}
Taking the limit as $k\rightarrow\pi$ gives a divergence of the alternating zigzag mode at the critical point 
\begin{equation}
\label{phimin}
\lambda_{c} = 2c.
\end{equation}
In the numerical simulations we shall consider an open chain instead of using periodic boundary conditions, and small finite-size corrections  lower slightly this value.

\section{II.\quad Full counting statistics of kinks as a function sampling}
\begin{figure}[h!]
\begin{center}
\includegraphics[width=1\linewidth]{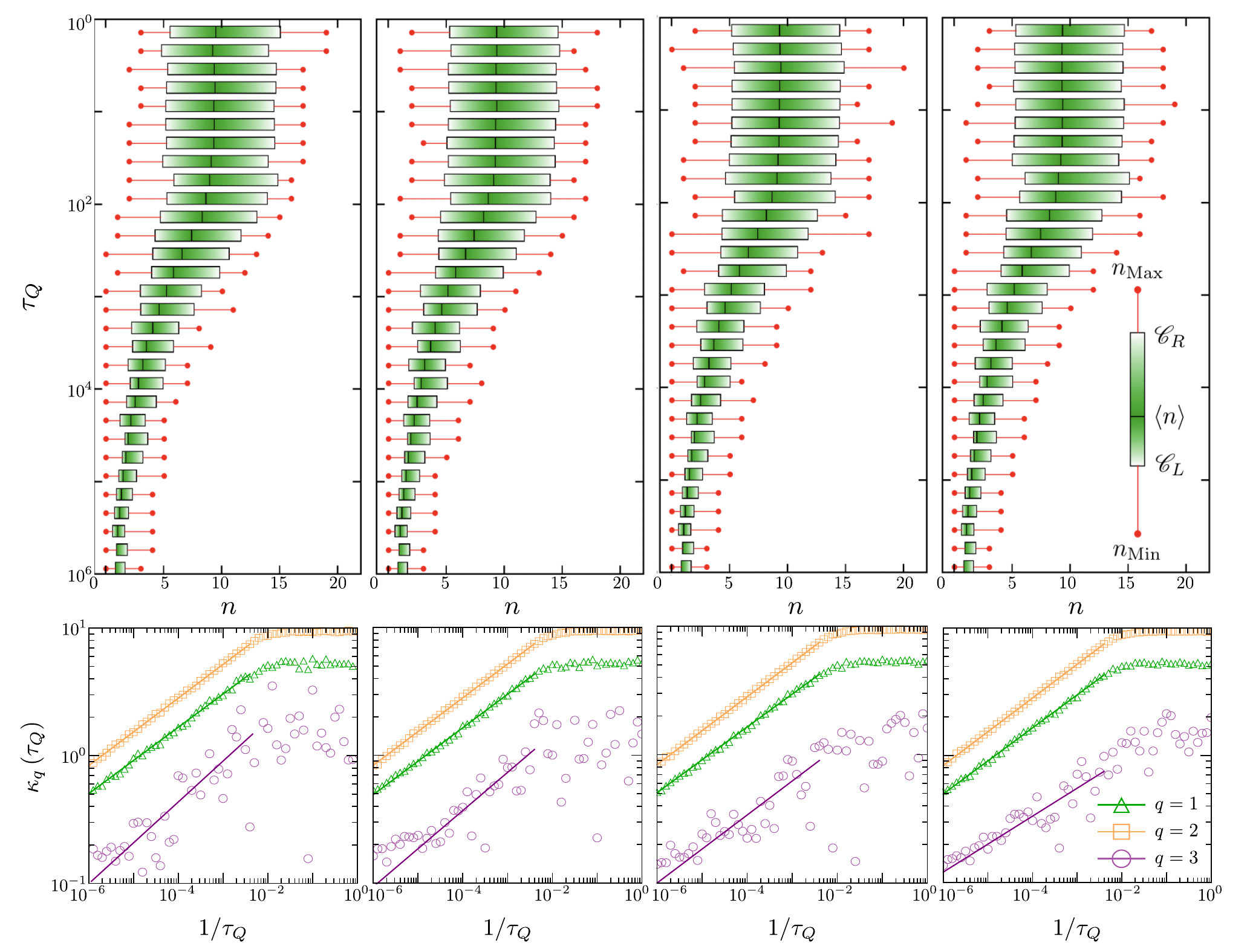}
\end{center}
\caption{\label{SM_fig_1}  {\bf Characterization of probability distribution of the number topological defects.} The upper panel shows the total distribution in a box-and-whisker chart, where $\mathscr{C}_{R}$ and $\mathscr{C}_{L}$ is given by Eq.~\eqref{SM_Col1}, and different quench time values $\tau_Q$ are considered for chain of  $N=100$ sites. The number of sampling trajectories is varied from 1000 (left) to 4000 (right).}
\end{figure}
In the main text,  we consider a one-dimensional chain exhibiting a structural phase transition between a linear and a doubly-degenerate zigzag  phase ($c>0$). We solve numerically the  Langevin dynamics described by the set of coupled stochastic differential equations
\begin{equation}
\ddot{\phi}_i+\eta \dot{\phi}_i+\partial_{\phi_i}\pap{\sum_j\frac{1}{2}[\lambda(t)\phi_j^2+\phi_j^4]+c\sum_j \phi_j \phi_{j+1}} +\zeta(t)=0, \quad i=1,\dots,N
\end{equation}
with constant friction $\eta>0$ and  nearest-neighbor coupling $c>0$ favoring the ferromagnetic order. Additionally, $\lambda(t)$ is ramped from a positive initial value to a negative one in a time scale $\tau_Q$ and $\zeta\pap{t}$ is a real Gaussian process with zero mean, satisfying $\la\zeta\pap{t}\zeta\pap{s}\ra=\sigma\delta(t-s)$. We make a swept in the number of realizations from 1000 to 4000, with $\lambda_0=1$, $\lambda_f=-1$, $c=1/2$, $\eta=50$, $\sigma=2\times 10^{3}$. For these parameters, under periodic boundary conditions $\lambda_c=2c=1$. For a linear chain, numerical simulations of the minimum energy configuration show that the critical point is actually slightly below this value, at $\lambda_c\simeq0.9995$ (the transition is actually slightly inhomogeneous as a result of the linear configuration). We have checked  that the results presented are robust against variations in the choice of these quench parameters. In particular, we have compared the numerics when starting the quench well above the critical point ($\lambda_0=2$) or close to it ($\lambda_0=1$, used throughout the manuscript), finding negligible differences, as can be expected from the symmetry of the ramp \cite{Antunes06}. The dynamics is over-damped with dynamic critical exponent $z=2$ and  $\nu=1/2$ \cite{SM_Laguna98,SM_delcampo10}. 
\\
In  Figure~\ref{SM_fig_1}, we characterize the full counting statistic of kinks as a function of the quench time $\tau_Q$ and the number of sampling trajectories considered. In the upper panels, we depict the behavior of the probability distribution in a box-and-whisker chart. The solid vertical line represents the mean number of kinks $\langle n\rangle$. The size of the left and right rectangles is fixed by
\begin{align}\label{SM_Col1}
\mathscr{C}_{R} =\sum_{n=n_{{\rm Min}}}^{\lceil \langle n \rangle \rceil -1}n\,P(n), \qquad \mathscr{C}_{L} =\sum_{n=\lceil \langle n \rangle \rceil}^{n_{{\rm Max}}}n\,P(n), 
\end{align}
where $n_{{\rm Min}}$ and $n_{{\rm Max}}$ are the number minimum and maximum of kinks obtained,  marked  with red points. In the low panel of Fig.~\ref{SM_fig_1}, we show the corresponding universal scaling of the cumulants $\kappa_q$ with $q\in\pac{1,2,3}$ and  report the values of the Kibble-Zurek exponent $\beta_{{\rm KZM}}$ obtained in the Table~\ref{SM_Tab_1}.     

\begin{table}[h!]
\centering
\begin{tabular}{| M{2cm}||  M{2.5cm} | M{1.5cm} || M{2.5cm} | M{1.5cm} ||  M{2.5cm}| M{1.5cm}|}\hline
 \multirow{2}{*} {\# Trajectories}& \multicolumn{2}{c||}{$\kappa_1 \propto \tau_Q^{-\beta_{{\rm KZM}}}$} &\multicolumn{2}{c||}{$\kappa_2 \propto \tau_Q^{-\beta_{{\rm KZM}}}$} &\multicolumn{2}{c|}{$\kappa_1 \propto \tau_Q^{-\beta_{{\rm KZM}}}$}\\\cline{2-7}  
 & $\beta_{{\rm KZM}}\pm\Delta \beta_{{\rm KZM}}$ & $r^2$ & $\beta_{{\rm KZM}}\pm\Delta \beta_{{\rm KZM}}$ &$r^2$ & $\beta_{{\rm KZM}}\pm\Delta \beta_{{\rm KZM}}$ &$r^{2}$\\\hline\hline
$1000$ & $0.264\pm 0.001$ & $0.9999$ & $0.260\pm0.005$ & $0.9978$ & $0.320\pm 0.072$ & $0.7221$\\\hline 
$2000$ & $0.264\pm 0.001$ & $0.9999$ & $0.262\pm0.002$ & $0.9993$ & $0.299 \pm 0.037$ & $0.9001$\\\hline
$3000$ & $0.264\pm 0.001$ & $0.9999$ & $0.256\pm0.003$ & $0.9993$ & $0.265\pm 0.036$ & $0.8878$ \\\hline
$4000$ & $0.264\pm 0.001$ & $0.9999$ & $0.254\pm0.002$ & $0.9991$ & $0.245\pm 0.026$ & $0.924$\\\hline
$5000$ & $0.261\pm 0.001$ & $0.9999$ & $0.257\pm0.001$ & $0.9997$ & $0.252\pm 0.019$ & $0.959$\\\hline
\end{tabular}
\caption{{\bf Numerical power-law exponents.} The cumulants of the  probability distribution for the number of topological defects generated across a continuous phase transition exhibit  a power-law  scaling $\kappa_{q} \propto\tau_Q^{-\beta_{\rm KZM}}$. Fitted power-law exponents $\beta_{\rm KZM}$ are shown  for different number of sampling trajectories and system size $N=100$.}\label{SM_Tab_1}
\end{table}
\begin{figure}[h!]
\begin{center}
\includegraphics[width=0.7\linewidth]{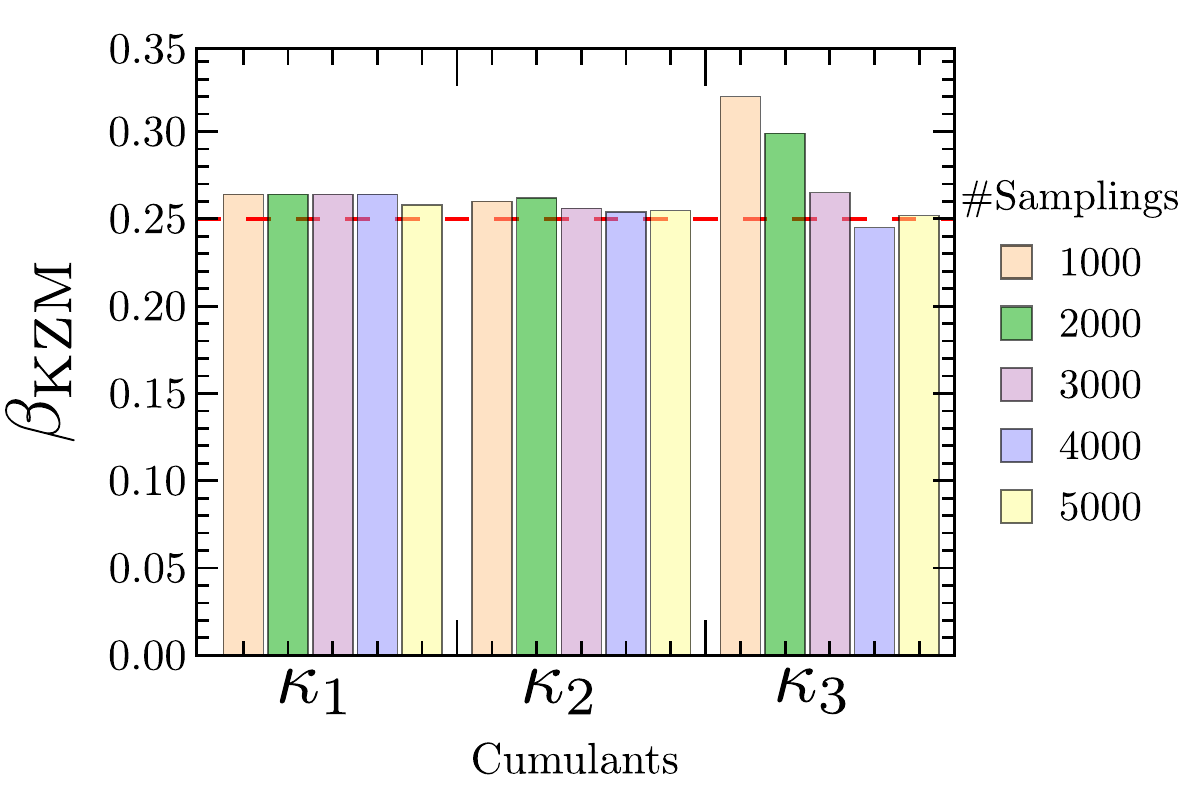}
\end{center}
\caption{\label{SM_Fig_2}  {\bf Test of convergence for the critical Kibble-Zurek exponent $\beta_{{\rm KZM }}$ as a function of number of trajectories.} The histogram shows the convergence on the power-law exponent governing the scaling of  each cumulants $\kappa_q$ ($q=1,2,3$) as the  number of sampling realizations is increased. The horizontal dashed line represents the Kibble-Zurek scaling exponent $\beta_{{\rm KZM}}=\nu/(1+z\nu)=1/4$ for $\nu=1/2$ and $z=2$.}
\end{figure}
Figure~\ref{SM_Fig_2} shows the comparison between the theoretical and numerical power-law exponents for varying sampling.  In addition, numerical simulations in Fig.~\ref{SM_Fig_2}  show the agreement between the theoretical an analytical power-law exponents for the first few cumulants.\\ 

\section{III.\quad Cumulant ratios}
\begin{figure}[h!]
\begin{center}
\includegraphics[width=1\linewidth]{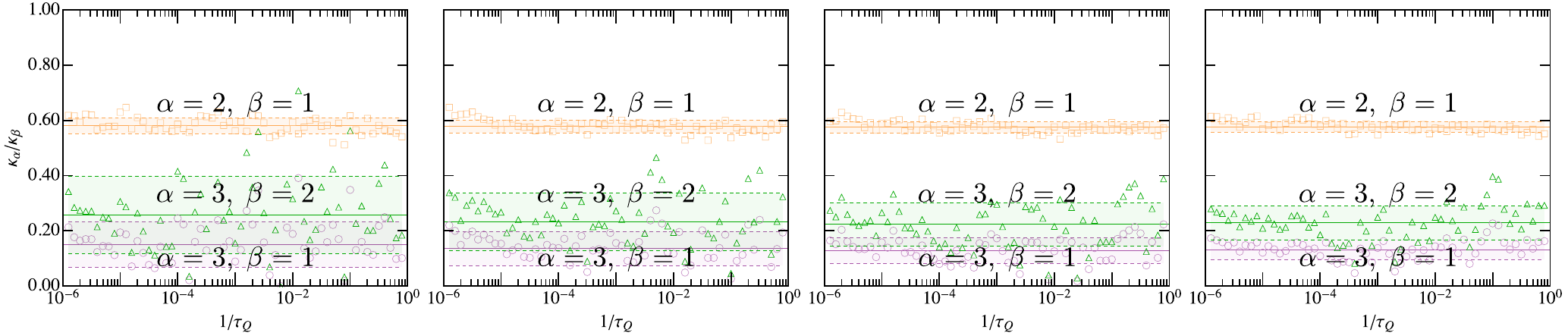}
\end{center}
\caption{\label{SM_fig_3} {\bf The ratios between $\kappa_\alpha$  and $\kappa_{\beta}$ cumulants.} Numerical results for the cumulant ratios ratio as a function of  the (inverse) quench time. The number of sampling trajectories is varied from 1000 (left) to 4000 (right). In all plots, the solid line is the mean value for the ratio $\kappa_{\alpha}/\kappa_{\beta}$ and the shadow region between two dashed lines corresponds to uncertainty associated with each ratio.}
\end{figure}
In the main text, we show that the ratio between any two cumulants is independent of the quench time and fixed by the probability $p$ for topological defect  formation at the location at which adjacent domains merge.  Figure~\ref{SM_fig_3} shows  the cumulant ratios $\kappa_{\alpha}/\kappa_{\beta}$ with $\alpha>\beta\in\{1,2,3\}$ for different number of sampling trajectories. The value of the ratios is constant (independent of the quench time) and uniquely fixed by the estimated probability for kink formation $p$, in agreement with the binomial distribution.  Naturally, as  the number of trajectories increases, the  uncertainty in the numerical value of the ratios is reduced, as shown by the shadowed region in  Fig.~\ref{SM_fig_3}. 
\begin{table}[h!]
\centering
\begin{tabular}{| M{2cm}||  M{2cm} | M{2cm} || M{2cm} | M{2cm} ||  M{2cm}| M{2cm}|}\hline
 \multirow{2}{*} {\# Trajectories}& \multicolumn{2}{c||}{$P_{1}=\kappa_2/\kappa_1$} & \multicolumn{2}{c||}{$P_{2}=\kappa_3/\kappa_1$} & \multicolumn{2}{c|}{$P_{3}=\kappa_3/\kappa_2$}\\\cline{2-7} 
 & $P_{1}$ & $\Delta P_1$ & $P_{2}$ & $\Delta P_2$ &$P_{3}$ & $\Delta P_3$\\\hline\hline
 $1000$ & $0.582$ & $0.029$ & $0.151$ & $0.082$ & $0.258$ & $0.139$\\\hline
 $2000$ & $0.579$ & $0.024$ & $0.136$ & $0.062$ & $0.233$ & $0.104$\\\hline
 $3000$ & $0.575$ & $0.019$ & $0.129$ & $0.046$ & $0.223$ & $0.079$\\\hline
 $4000$ & $0.578$ & $0.019$ & $0.132$ & $0.035$ & $0.229$ & $0.061$\\\hline
 $5000$ & $0.578$ & $0.015$ & $0.135$ & $0.035$ & $0.234$ & $0.061$\\\hline\hline
\end{tabular}
\caption{{\bf Cumulant ratios.} Numerical results for the mean cumulant ratios  as a function of  the number of trajectories.}\label{SM_Tab_2}
\end{table}

In the Table.~\ref{SM_Tab_2}, we report the average value of the cumulant ratios  and the corresponding uncertainty.\\  

\subsection{A.\quad Numerical estimation of $p$}
\begin{figure}[h!]
\begin{center}
\includegraphics[width=1\linewidth]{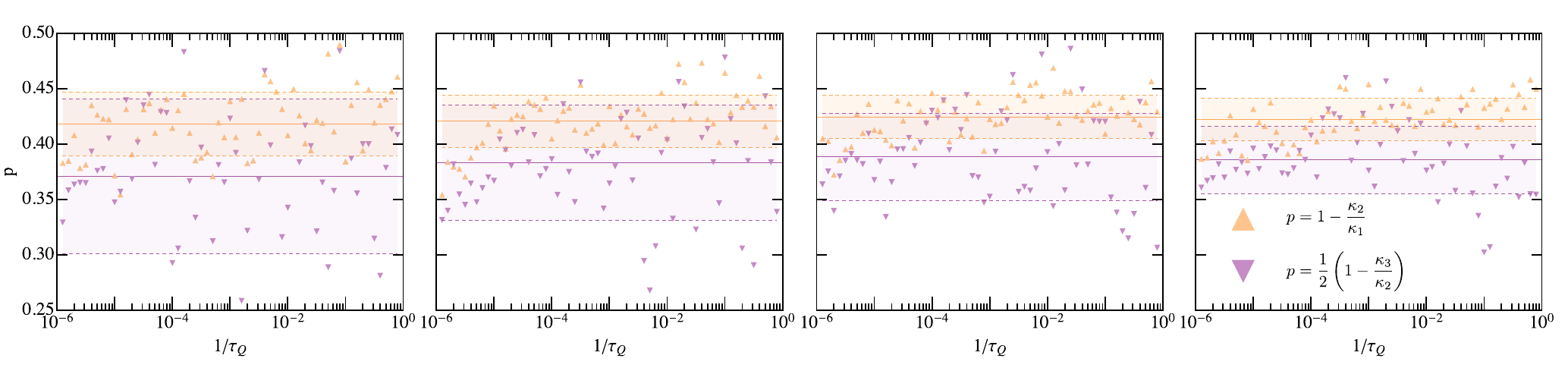}
\end{center}
\caption{\label{SM_fig_4} {\bf Numerical estimation of $p$ as a function of quench time.} For the up and down triangles symbols, we consider that the $p$ value is given by Eq.~\eqref{rel_p1} and Eq.~\eqref{rel_p3}, respectively. The number of trajectories is changed from 1000 (left) to 4000 (right) for a chain of  $N=100$ sites. In all plots, the solid line is the averaged estimated value of  $p$  and the shadowed region between two dashed lines reflects the corresponding uncertainty.}
\end{figure}

In the main text, we show how the cumulant ratios are fixed by the probability $p$ in a Bernouilli trial according to
\begin{align}
\frac{\kappa_2}{\kappa_1} &= 1 - p,\label{rel_p1}\\
\frac{\kappa_3}{\kappa_1} &=\pap{1-p}\pap{1-2p}.\label{rel_p2}
\end{align}
By direct substitution of Eq.~\eqref{rel_p1} into Eq.~\eqref{rel_p2}, we obtain that $p$ satisfy:
\begin{equation}\label{rel_p3}
p=\frac{1}{2}\pap{1-\frac{\kappa_3}{\kappa_2}}.        
\end{equation}
In the Figure~\ref{SM_fig_3}, we show that ratios $ \kappa_\alpha / \kappa_\beta $ with $\alpha > \beta$ and $\alpha,\beta\in\pac{1,2,3}$ are constant and independent of the quench time. In this way, we assumed that every ratio has an uncertain constant given by $\Delta P_\alpha$ where $\alpha\in\pac{1,2,3}$, following the notation of Table~\ref{SM_Tab_2}. Therefore,  the constant $p$ satisfies
\begin{align}
p&=\frac{1}{2}\pap{1-P_3}\pm \Delta_{P_3} p, & p&= \pap{1 - P_1}\pm \Delta_{P_1} p,\\
p&=p_{{\rm TEO}}\pap{P_3} \pm \Delta_{P_3} p, & p&= p_{{\rm TEO}}\pap{P_1}\pm \Delta_{P_1} p,
\end{align}    
where $P_1 = \kappa_2 /  \kappa_1$, $P_3 = \kappa_3 /  \kappa_2$, $\Delta_{P_3} p$ and $\Delta_{P_1} p$ are the corresponding uncertainty. Using, standard error propagation, it follows that
\begin{align}
\Delta_{P_1}^{2} p&=\left|\frac{\partial}{\partial P_1} p_{{\rm TEO}}\pap{P_1}\right|^{2}\pap{\Delta P_1}^{2}\Rightarrow \Delta_{P_1}p =\Delta P_1,\\
\Delta_{P_3}^{2} p &=\left|\frac{\partial}{\partial P_3}p_{{\rm TEO}}\pap{P_3}\right|^{2}\pap{\Delta P_3}^{2}\Rightarrow \Delta_{P_3}p =\frac{1}{2}\Delta P_3 .
\end{align} 
In  Figure~\ref{SM_fig_4}, we depict the numerical estimated value of $p$ using the relations obtained in Eq.~\eqref{rel_p1} and Eq.~\eqref{rel_p3} as a function of quench time. The number of trajectories is swept from 1000 to 5000, and we report the estimated  $p$  in Table~\ref{SM_Tab_3}. We  note that $p$ is approximately constant as a function of quench time. The estimated value  is obtained by seeking convergence as the number of trajectories is  increased.  
       
\begin{table}[h!]
\centering
\begin{tabular}{| M{2cm}||  M{2cm} | M{2cm} || M{2cm} | M{2cm} |}\hline
 \multirow{2}{*} {trajectories}& \multicolumn{2}{c||}{$p_{{\rm TEO}}\pap{P_1}=1-P_1$} & \multicolumn{2}{c|}{$p_{{\rm TEO}}\pap{P_3}=\frac{1}{2}\pap{1-P_3}$} \\\cline{2-5} 
 & $p_{{\rm TEO}}\pap{P_{1}}$ & $\Delta_{P_1}p$ & $p_{{\rm TEO}}\pap{P_{3}}$ & $\Delta_{P_3}p$ \\\hline\hline
 $1000$ & $0.418$ & $0.029$ & $0.371$ & $0.070$\\\hline
 $2000$ & $0.421$ & $0.024$ & $0.383$ & $0.052$\\\hline
 $3000$ & $0.425$ & $0.019$ & $0.388$ & $0.039$\\\hline
 $4000$ & $0.422$ & $0.019$ & $0.386$ & $0.031$\\\hline
 $5000$ & $0.422$ & $0.015$ & $0.383$ & $0.030$\\\hline
\end{tabular}
\caption{{\bf Numerical estimation of $p$ as a function of  sampling.}}\label{SM_Tab_3}
\end{table}
\newpage
\section{IV.\quad Onset of  adiabaticity}
\begin{figure}[h!]
\begin{center}
\includegraphics[width=1\linewidth]{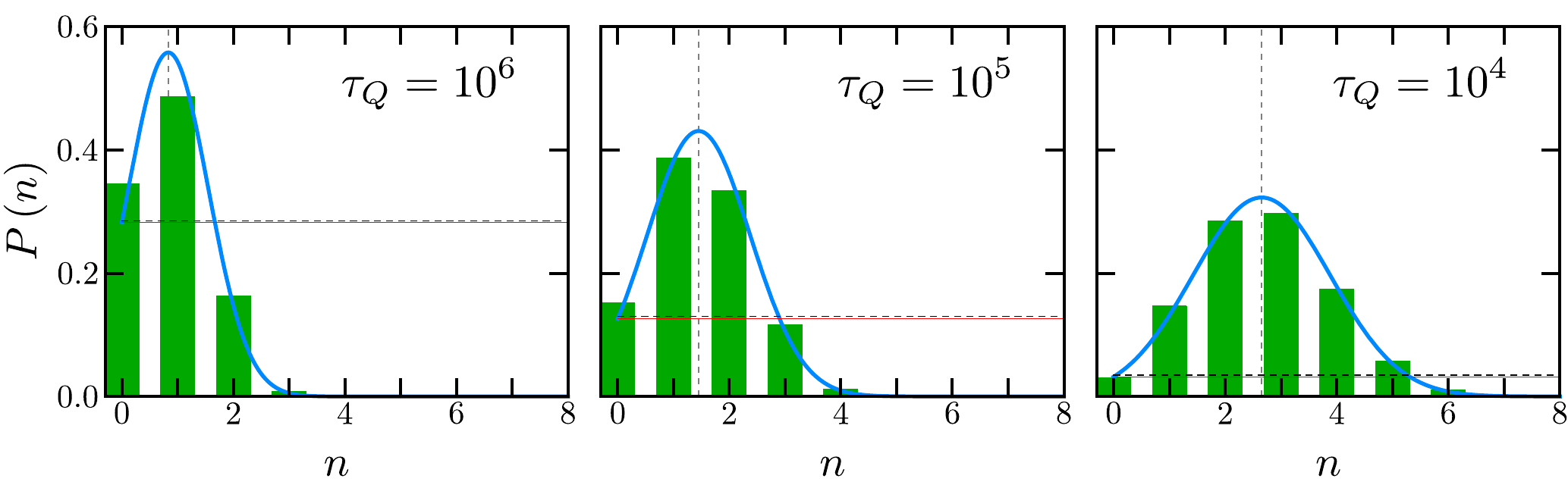}
\end{center}
\caption{\label{SM_fig_5} {\bf Onset of Adiabaticity.} Probability distribution of the number of kinks $P(n)$ generated with  a long time quench $\tau_Q$. The  distribution depicted by a solid line shows the agreement between the theoretical (Eq.12 in the main text) and numerical results. Additionally, the horizontal solid ($p$ given by Eq.~\eqref{rel_p1}) and  dashed ($p$ given by Eq.~\eqref{rel_p3}) lines show the probability for no kinks $P(0)$. The probability distribution is built by sampling over an ensemble of 15000 trajectories.}
\end{figure}

As shown in the main text,  the probability for zero defects decays exponentially with the mean number of defects $P(0)=\exp\pas{-\la n\ra}$, and is given by 
\begin{equation}
P(0)=\frac{1}{\sqrt{2\pi(1-p)\la n\ra}}\exp\pas{-\frac{1}{2\pap{1-p}}\la n\ra},
\end{equation}
whenever the normal approximation to the distribution can be invoked.
The mean value is dictated by the KZM. The actual distribution 
becomes  manifestly non-symmetric around the mean value at the
 onset of adiabaticity, when $P(0)$ is significant. In this limit, the normal approximation ceases to be accurate, and so it does the expression for $P(0)$ derived form it. The accuracy of the normal approximation is recovered for faster quenches, as shown in Fig. (\ref{SM_fig_5}). 

\section{V.\quad Tails of the number distribution of topological defects}
\begin{figure}[h!]
\begin{center}
\includegraphics[width=1.01\linewidth]{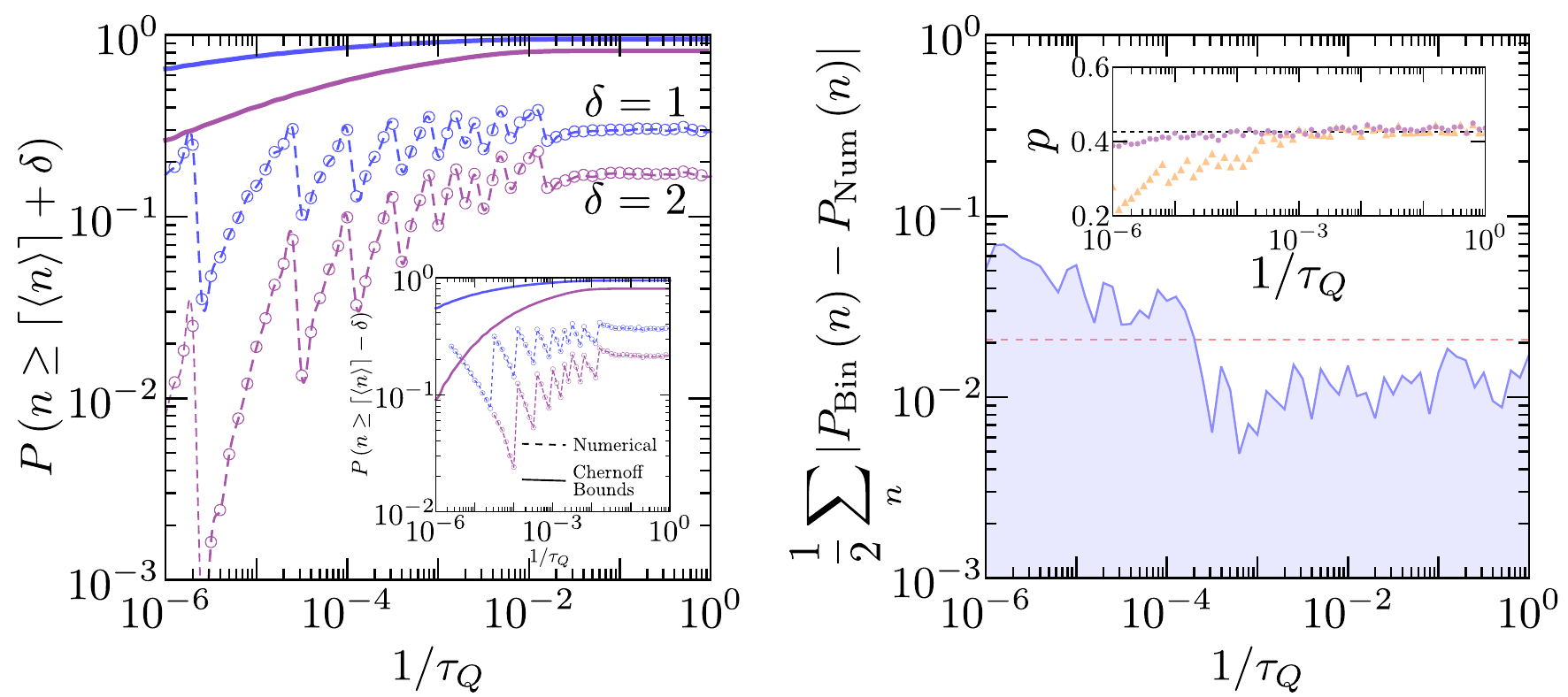}
\end{center}
\caption{\label{SM_fig_6} {\bf Tails of the number distribution of topological defects.} {\it Left:} In the main and insert panel, we show the Chernoff bound for the upper and lower (solid line) tails of the distribution and show the corresponding numerical cumulative probability (symbols - dashed line). The results for $\delta = 1$ and $\delta = 2$ are shown in the  blue and purple color,  respectively. {\it Right:} The main panel shows  the relative error between the binomial and numerical distribution, the dashed red line correspond at average error. In the inset, we show the $p$  value estimated in two different ways, based on Eq.~\eqref{rel_p1} ($\bullet$), and the traditional distribution fit method ``maximize the log-likelihood function" ($\blacktriangle$).  The number of trajectories is 15000.}
\end{figure}

Knowledge of the distribution of topological defects beyond the KZM raises the question as to the width of the distribution and the probability of having large deviations from the mean value. 

In the main text, we have analyzed the probability of occurrence of zero kinks in the final nonequilibrium state after crossing the phase transition. This requirement of adiabaticity may however be too strict and relaxed notions can be imposed by  bounding the tail of the distribution associated with high kink numbers.
To this one can consider, general bounds on the tails of the distribution or the exact computation of the cumulative probability associated with these tails.
 
In the first case, we can use the Chernoff bound according to which the lower and upper tails of the distribution are constrained by the inequalities. 
\begin{align}
{\rm Prob}(n\leq \lceil\la n\ra\rceil-\delta)&\leq e^{-\frac{\delta^2}{2\la n\ra}}&\text{(Lower tail)}, &&{\rm Prob}(n\geq \lceil \la n\ra\rceil +\delta)&\leq e^{-\frac{\delta^2}{2\pap{\la n\ra +\delta/3}}}&\text{(Upper tail)}.
\end{align}

The relevance of these bounds is shown in Fig, \ref{SM_fig_6}, for different values of $\delta\in\mathbb{N}$ and as a function of the 
quench time. Different panels correspond to increasing number of trajectories, from left to right. Large deviation theory can be used to bound these events.

Pursuing the second  approach, we resort to the direct computation of the cumulative probability associated with large deviations.
First, we consider the binomial distribution that exactly describes the distribution of the number of topological defects according to our model, and for which
\beqa
\label{etails}
 {\rm Prob}(n\leq k)=I(1-p;\mathcal{N}-k,k+1),\quad {\rm Prob}(n>k)=I(p;k+1,\mathcal{N}-k),
 \eeqa
in terms of the regularized beta function 
\beqa
I(x;a,b)=\frac{B(x;a,b)}{B(a,b)}=\frac{\int_0^xy^{a-1}(1-y)^{b-1}dy}{\int_0^1y^{a-1}(1-y)^{b-1}dy},
\eeqa
where the incomplete and complete beta functions are denoted by $B(x;a,b)$ and $B(a,b)$, respectively. Deviations away from the mean can be accounted for by taking, e.g., $k=\lceil\la n\ra\rceil\pm\delta$.
While exact, these expressions do not exhibit clearly the dependence on the quench time, that is encoded in 
the value of 
\beqa
\mathcal{N}=\frac{{\rm Vol}}{f\xi_0^D}\left(\frac{\tau_0}{\tau_Q}\right)^{\frac{D\nu}{1+z\nu}},
\eeqa
which should be taken to be an integer (e.g., the floor function of the right hand side).

To bring out the dependence on $\tau_Q$,  we further consider deviations away form the mean in the normal approximation for which
\beqa
{\rm Prob}(|n-\la n\ra|> \delta)=\int_{\la n\ra +\delta}^\infty dy\frac{1}{\sqrt{2\pi{\rm Var}(n)}}e^{-\frac{(y-\la n\ra)^2}{2{\rm Var}(n)}}=\frac{1}{2}{\rm Erfc}\left(\frac{\delta}{2{\rm Var}(n)}\right),
\eeqa
where ${\rm Erfc}(x)=\frac{2}{\sqrt{\pi}}\int_0^xe^{-y^2}dy$ is the complementary error function.

In the context of KZM, the argument explicitly reads
\beqa
x=\frac{\delta}{2{\rm Var}(n)}=\frac{\delta f\xi_0^D}{2(1-p)p{\rm Vol}}\left(\frac{\tau_Q}{\tau_0}\right)^{\frac{D\nu}{1+z\nu}}.
\eeqa
The case of  small deviations from the mean and/or moderate driving correspond to $x\ll 1$. 
To leading order in  $x$ and recalling that ${\rm Var}(n)=(1-p)\la n\ra$, we find
\beqa
\label{nearmean}
{\rm Prob}(|n-\la n\ra|> \delta)&=&\frac{1}{2}-\frac{\delta}{\sqrt{2\pi(1-p)\la n\ra }}\\
&=&\frac{1}{2}-\delta\left(\frac{f\xi_0^D}{2\pi(1-p)p{\rm Vol}}\right)^{\frac{1}{2}}\left(\frac{\tau_Q}{\tau_0}\right)^{\frac{D\nu}{2(1+z\nu)}}.
\eeqa
Thus, the probability for deviations away of the mean decreases from half unit value with a universal power-law of the quench rate.

The opposite extreme corresponds to slow quenches within the validity of the normal approximation or rare events associated with large deviations in the sense that $x=\delta/2{\rm Var}(n)\gg 1$.
 Then, taking the leading term in the corresponding asymptotic expansion
 \beqa
{\rm Erfc}(x)=\frac{e^{-x^2}}{x\sqrt{\pi}}\left[1+\sum_{n=1}^\infty\frac{(2n-1)!!}{(2x^2)^n}\right],
\eeqa
one finds
\beqa
\label{farmean}
{\rm Prob}(|n-\la n\ra|> \delta)&=&\frac{(1-p)\la n\ra}{\delta \sqrt{\pi}}\exp\left(-\frac{\delta^2}{4(1-p)^2\la n\ra^2}\right)\\
&=&\frac{p(1-p){\rm Vol}}{\delta\sqrt{\pi}f\xi_0^D}\left(\frac{\tau_0}{\tau_Q}\right)^{\frac{D\nu}{1+z\nu}}
\exp\left[-\left(\frac{\delta f \xi_0^D}{2p(1-p){\rm Vol}}\right)^2\left(\frac{\tau_Q}{\tau_0}\right)^{\frac{2D\nu}{1+z\nu}}\right],
\eeqa
where we have emphasized the universal dependence on the quench time.
By contrast to the  Eqs.~\eqref{etails} that are exact for the binomial model, expressions (\ref{nearmean}) and (\ref{farmean}) are naturally restricted to the validity of the normal approximation for $P(n)$.

\newpage

\end{document}